\documentclass[twocolumn,preprintnumbers,amsmath,pra,showpacs]{revtex4}

\usepackage{graphicx}

\usepackage{dcolumn}
\usepackage{bm}

\def\gappeq{\mathrel{ \rlap{\raise.5ex\hbox{$>$}}
                      {\lower.5ex\hbox{$\sim$}} } }
\def\lappeq{\mathrel{ \rlap{\raise.5ex\hbox{$<$}}
                      {\lower.5ex\hbox{$\sim$}} } }

\begin{document}

\preprint{PRA}

\title{Coherent cross-talk and parametric driving of matter-wave vortices}
\author{N. G. Parker}
\email{nick.parker@ncl.ac.uk}
\author{A. J. Allen}
\author{C. F. Barenghi}
\author{N. P. Proukakis}
\affiliation{School of Mathematics and Statistics, Newcastle University, Newcastle upon Tyne, NE1 7RU, UK}
\date{\today}
\begin{abstract}
We show that the interaction between vortices and sound waves in atomic Bose-Einstein condensates can be elucidated in a double-well trap:
with one vortex in each well, the sound emitted by each precessing vortex can be driven into the opposing vortex (if of the same polarity).  This cross-talk leads to a periodic exchange of energy between the vortices which is long-range and highly efficient.  The increase in vortex energy (obtained by numerical simulations of the Gross-Pitaevskii equation) is significant and experimentally observable as a migration of the vortex to higher density over just a few precession periods.   Similar effects can be controllably engineered by introducing a precessing localised obstacle into one well as an artificial generator of sound, thereby demonstrating the parametric driving of energy into a vortex.

\end{abstract}
\maketitle

\section{Introduction}
In a quantum fluid, such as an atomic Bose-Einstein condensate (BEC)
or superfluid Helium, vortices possess quantized circulation,
synonymous with them being a topological defect in the macroscopic
condensate phase. Quantized vortex lines, rings, lattices and
tangles have been the subject of experimental study in superfluid
Helium for over 50 years \cite{Donnelly1991}, and in which recent
emphasis has been on their role in quantum turbulence
\cite{Barenghi2001,Vinen2010}. Meanwhile, since the late 1990's,
there has been fast-growing interest in vortices in Bose-Einstein
condensates \cite{Fetter2001,Anderson2010}, where the
controllability of these gases has led to the generation of single
vortices \cite{Matthews1999,Rosenbusch2002}, giant vortices
\cite{Engels2003}, vortex dipoles \cite{Neely2010}, soliton-vortex
hybrids \cite{Ginsberg2005} and turbulent vortex tangles
\cite{Henn2009}.  It is worthy of note that recent breakthroughs in
imaging of both Helium \cite{Bewley2006} and BEC \cite{Freilich2010}
systems now enable the dynamics of quantized vortices to be
monitored in real-time.

The full nature of the vortex-sound interaction in quantum fluids is
unclear \cite{Buhler2010}.  The superfluid topology constrains
quantized vortices to disappear by annihilating with an oppositely
charged vortex or vanishing at the edge of the system (where they
annihilate with their image). In the limit of zero temperature and
for a uniform condensate, sound waves are the low-lying excitations
of the system and provide the only energy sink for vortex decay. For
example, at zero temperature the reconnection of vortex lines
\cite{Leadbeater2001} and the acceleration of a vortex line segment
both generate sound waves \cite{Lundh2000,Vinen2001}, dissipating
the vortical energy.  In the latter case, the acceleration that
drives sound emission may arise from the influence of other vortices
\cite{Proukakis2004,Barenghi2005}, Kelvin waves excitations of
vortex lines \cite{Leadbeater2003} or the Magnus force in an
inhomogeneous ambient density, e.g. in a trapped condensate
\cite{Parker2004a}.  The experimentally observed decay of vorticity
at very low temperature in superfluid He \cite{Davis2000,
Walmsley2007} is thought to be primarily due to the Kelvin waves
dissipation route, with reconnections playing a secondary role
\cite{Vinen2010}.

Less well understood is the inverse process, i.e. the absorption of
sound by a vortex.  Insight may be gleaned from vortex-sound
interactions in fluid dynamics \cite{Buhler2010}.  For example, an
acoustic ray model has predicted that certain trajectories of sound
wave can spiral into the vortex core, transferring energy to the
vortical flow \cite{Nazarenko1994,Nazarenko1995}.  While sound waves
can induce the nucleation of vortices through the collapse of
cavitating bubbles \cite{Schwarz1981,Barenghi2004}, sound absorption
by pre-existing vortices is not thought to play a significant role
in homogeneous superfluid Helium systems.  However, sound absorption
may become considerable in atomic BECs due to their confined
geometry. Indeed, the lack of sound-induced decay of a vortex
precessing in a harmonically-trapped BEC has been attributed to
reabsorption of the emitted sound \cite{Parker2004a} (although
related works \cite{Lundh2000,Fetter2001} predict that the sound
emission may be prohibited due to the sound wavelength exceeding the
system size).   The harmonic nature of the trap appears key to
supporting a sound-vortex equilibrium, with trap anharmonicities
apparently leading to net vortex decay
\cite{Parker2004a,Kevrekidis2003}, in analogy to dark solitons
\cite{Busch2000,Parker2010}. However, it is difficult to resolve the
interaction of sound and vortices in single trapped condensates
\cite{Parker2004a,Barenghi2005} due to their co-habitation in the
trap.

Analogous questions exist over the interaction of matter-wave dark
solitons with sound.  Like vortices, they radiate sound waves under
acceleration \cite{Busch2000,Parker2003}, become stabilised in
harmonic traps \cite{Parker2004a} and can be parametrically driven
by artificially generated sound waves \cite{Proukakis2004}.
Recently, a double-well trap was shown to offer beneficial insight
into the soliton-sound interaction \cite{Allen2011}.  With one
soliton in each well, large-scale exchanges in energy between the
solitons clearly demonstrated a long-range sound-mediated
interaction.

Here we will exploit a double well trap to study the emission and
absorption of sound by vortices.  We will begin by considering an
idealized double trap geometry.  The trap system and theoretical
model are outlined in Section II.  In Section III we consider how a
single vortex behaves in this system (while not the main results of
our work, this an essential prerequisite to understanding the
dynamics in later sections).  In Section IV we progress to consider
how two vortices, one in each well, interact via the exchange of
sound waves or ``cross-talk".  In Section V we replace one of the
vortices with a moving obstacle, and explore how the sound generated
interacts with the remaining vortex.  In Section VI we discuss our
theoretical findings.  In Section VII we demonstrate the same
qualitative behaviour in an experimentally-achievable double trap
geometry.  Finally, in Section VIII we draw conclusions of our work.

\section{Theoretical framework and trap set-up}

We consider a BEC at ultracold temperatures such that thermal and
quantum fluctuations can be neglected and that the system is well
parameterised by a mean-field order parameter $\Psi({\bf r},t)$
which satisfies the three-dimensional Gross-Pitaevskii equation
(GPE) \cite{Pethick2002}.  We assume a quasi-2D geometry, whereby
harmonic trapping is sufficiently tight in one dimension, taken here
to be the $z$-direction, to freeze out the corresponding dynamics.
Then, using the decomposition $\Psi({\bf
r},t)=\psi(x,y,t)\psi_{z}(z)$, one can integrate out the
time-independent axial component $\psi_z(z)$ from the 3D GPE.  The
transverse order parameter $\psi(x,y,t)$ then satisfies the 2D GPE,
\begin{equation}
i\hbar \partial_t \psi=\left(-\frac{\hbar^2}{2m}\nabla^2+V(x,y)+g |\psi|^2 -\mu\right)\psi,
\end{equation}
where $V(x,y)$ is the axial trapping potential and $m$ is the atomic mass.  The 2D chemical potential $\mu$ is related to the 3D chemical potential $\mu'$ via $\mu=\mu'-\hbar \omega_z/2$, where $\omega_z$ is the trap frequency in the axial direction.   {\it s}-wave atomic interactions, of length $a$, give rise to the nonlinear term with coefficient $g=2\sqrt{2\pi}\hbar^2 a/ml_z$, where $l_z=\sqrt{\hbar/m\omega_z}$ is the harmonic oscillator length of the frozen dimension.

 The complex order parameter $\psi(x,y,t)$ can be written as $\psi(x,y,t)=\sqrt{n(x,y,t)}\exp[i\phi(x,y,t)]$, where $n(x,y,t)$ and $\phi(x,y,t)$ are the distributions of atomic density and phase, respectively.   Furthermore, the phase defines the fluid velocity ${\bf v}=(\hbar/m)\nabla \phi$.
In 2D vortices are singular points about which the phase wraps around by an integer multiple of $2\pi$ and the condensate flows azimuthally.  The density is pinned to zero at the central point creating a well-defined vortex core which relaxes to its unperturbed value at a distance of the order of the healing length $\xi=\hbar/\sqrt{m n g}$.

\begin{figure}[t]
\centering
\includegraphics[width=0.87\columnwidth,clip=true]{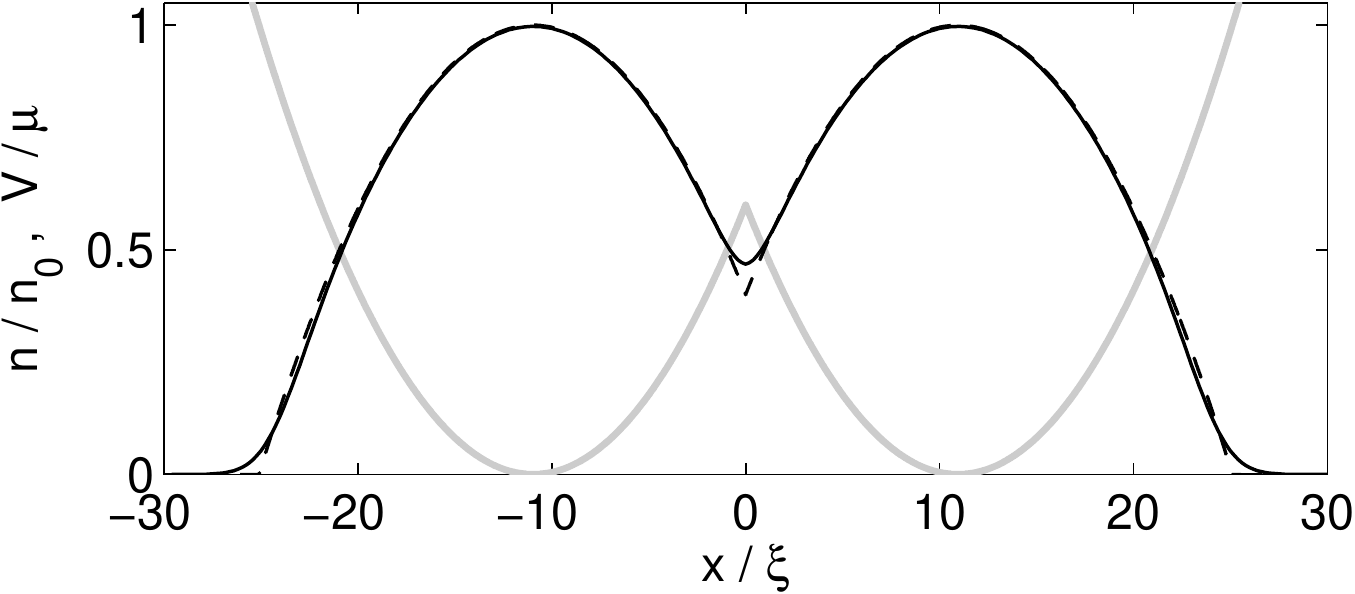}
\caption{Vortex-free density profile at $y=0$ along the $x$-axis (black solid line) in the double harmonic potential (grey solid line) with a barrier height at $x=0$ of $V_0=0.6\mu$ .  The corresponding Thomas-Fermi profile $n_{\rm TF}=(\mu-V)/g$ is also shown (black dashed line).}
\label{fig:trap}
\end{figure}

We will initially consider an idealized double well system consisting of two connected harmonic traps,
\begin{equation}
V(x,y)=\frac{1}{2}m\omega^2\left[\left(|x|-x_c \right)^2+y^2 \right].
\label{eqn:double_harmonic_potential}
\end{equation}
Each trap is circularly symmetric with frequency $\omega$ and displaced from the origin by $\pm x_c$, as illustrated in Fig. \ref{fig:trap}. The barrier separating the trap, which has a minimum height of $V_0=\frac{1}{2}m\omega^2 x_c^2$, determines the connectivity between the wells and thus the degree to which sound and vortices can propagate between wells.  The transfer of sound waves, which have energy of order $\mu$, between the wells will be possible for $V_0<\mu$ and prohibited for $V_0\gg \mu$.  The capacity for vortices to propagate between wells will depend additionally on the energy of the vortex.

While this trap is not directly achievable in experiments due to its
sharp feature at $x=0$, it is convenient to consider theoretically:
it identifies the main physical effect in a ``clean" manner and
allows us to draw on the established knowledge for vortices in
harmonic traps. We will later demonstrate that the same dynamics
persist in experimentally-achievable set-ups.

The 2D GPE is solved numerically using the Crank-Nicholson method
\cite{Minguzzi2004}. Over a typical simulation, the relative change
in norm and energy $\Delta N/N$ and $\Delta E/E$ are of order
$10^{-6}$, i.e. the solution is numerically well converged. The
vortex-free ground state, with density profile $n_{\rm VF}(x,y)$, is
found by propagating the 2D GPE in imaginary time. Vortical states
are imposed by forcing the phase distribution (during imaginary time
propagation) to,
\begin{equation}
\phi(x,y)=\prod_i q_i {\rm arctan} \left(\frac{y-y_i}{x-x_i}\right),
\end{equation}
where $i$ is the index of a vortex with charge $q_i$ located at
$(x_i,y_i)$.   Since multiply-charged vortices are energetically
unfavourable compared to multiple singly-charged vortices
\cite{Donnelly1991,Pethick2002}, we shall only consider
singly-charged vortices $|q_i|=1$ (relative polarity may change). We
will infer the change in energy of vortices in our system through
changes in their position, e.g. a drift to lower density signifies a
decrease in the vortex energy.  This is exactly what would be done
in reality, since the vortex energy cannot be directly measured
experimentally.

We assume units in which length, speed and energy are expressed in terms of the 2D healing length $\xi=\hbar/\sqrt{m n_0 g}$, speed of sound $c=\sqrt{n_0 g/m}$ and chemical potential $\mu=n_0 g$, where $n_0$ is the peak density.

The ratio $\mu/\hbar \omega$ specifies the nature of the condensate.
For $\mu/\hbar \omega\ll1$ the trap dominates and the ground state
will approximate the gaussian harmonic oscillator.  For $\mu/ \hbar
\omega \gg 1$ the interactions dominate and lead to a broad
condensate profile.  Then the kinetic energy of the ground state,
which depends on the gradient of the density, becomes negligibly
small. Under this Thomas-Fermi (TF) approximation, the density takes
the analytic form $n_{\rm TF}=(\mu-V)/g$ \cite{Pethick2002}.

We focus on a system with $\mu/\hbar \omega=10$. An example density
profile is shown in Fig.~\ref{fig:trap}.  It is closely matched by
the TF prediction, with the only significant deviation arising at
the condensate perimeter and the barrier.  There the density
gradient is not negligible, resulting in a smoothening of the
density over a lengthscale $\sim \xi$. The TF approximation predicts
a condensate radius $R_{\rm TF}=\sqrt{2\mu/m\omega^2}=14.14\xi$.
While we hereafter express length in terms of healing length, this
can be trivially related to the trap harmonic oscillator length
$l_{\rm ho}=\sqrt{\hbar/m\omega}$ via $l_{\rm ho}=\sqrt{10}\xi$.

\section{Single vortex}
We first explore the dynamics of a single vortex within the double
trap system (\ref{eqn:double_harmonic_potential}).  A vortex tends
to follow a path of equipotential through a trapped condensate,
e.g., precessing around the trap centre in a single harmonic trap
\cite{Rokhsar1997}.  In the double trap we can additionally
anticipate a regime in which the vortex can follow a dumbbell-shaped
path around both wells.  As is well known for single harmonic traps,
the energy and angular momentum associated with the vortex increases
as the vortex is moved to higher density, i.e. towards the centre of
the well \cite{Fetter2001,Pethick2002}.

We place a vortex in the right-hand well at a position
$(x_c,y_c+r_{v})$, where $(x_c,y_c)=([2V_0/m\omega^2]^{1/2},0)$ is
the origin of the right-hand trap and $r_{v}$ is the initial offset
of the vortex from the well centre.  NB the dynamics are insensitive
to the direction of the off-set.  The ensuing dynamics are sensitive
to the size of the vortex displacement $r_v$ and the inter-trap
barrier $V_0$.  We numerically evolve the vortex dynamics over a
long simulation time ($5000~(\xi/c)$) within this parameter space.
Note when the vortex is initially positioned very close to BEC edge
(typically within one healing length of $R_{\rm TF}$) its evolution
is indistinguishable from surface excitations that are generated.
This is a general consequence of being initiated in the low density
periphery and occurs even in isolated harmonic traps. As such we do
not present the vortex dynamics at such extreme positions.

\begin{figure}[b]
\centering
\includegraphics[width=0.8\columnwidth,clip=true]{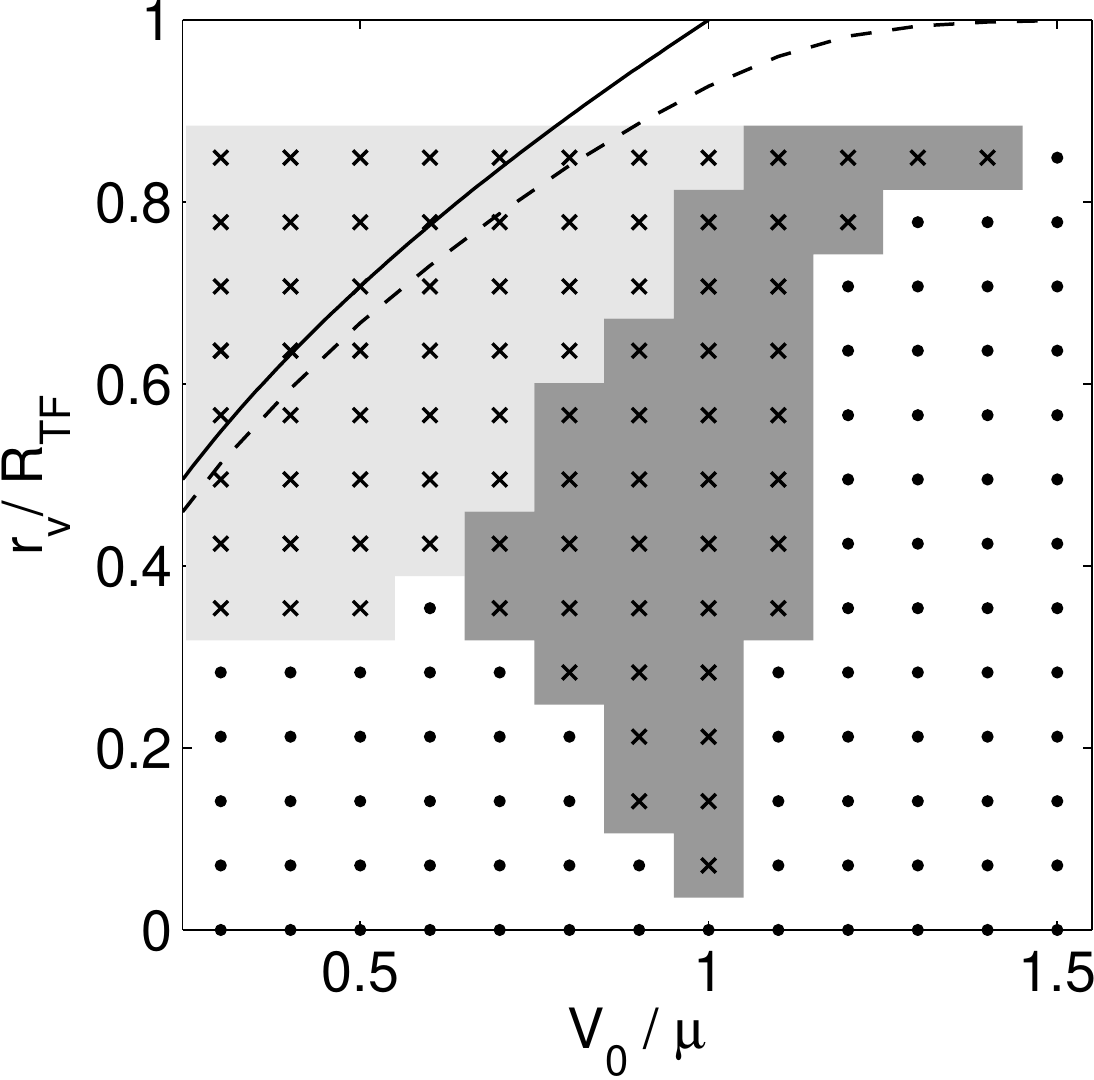}
\caption{Phase diagram in $r_v-V_0$ space for the single vortex case showing whether the final state of the system (after a long simulation time of $5 \times 10^3~(\xi/c)$) is a single vortex (dots) or the vortex-free state (crosses).  The light-shaded region is the cross-over regime (case I), in which the vortex passes into the far well.  The dark-shaded region is the inductive regime (case II), in which the initial vortex induces other vortices.   The lines are the TF predictions for the onset of cross-over dynamics $r_v/R_{\rm TF}=\sqrt(1-n_{\rm min}/n_0)$ using the TF prediction (solid  line) and the numerical values (dashed line) for $n_{\rm min}$.
The observed deviation between these lines and the light/dark grey boundary acts as an indication of the importance of sound emission in the actual vortex dynamics.
The TF radius of each well is $R_{\rm TF}=14.14\xi$.}
\label{fig:phase_diagram_single_vortex}
\end{figure}

The phase diagram for the vortex dynamics is plotted in
Fig.~\ref{fig:phase_diagram_single_vortex}, separating regions of
`stable' vortex dynamics (dots) from those where there is no vortex
within the system in the `final' simulated state (crosses), as
discussed in detail below. Note that we observe a qualitatively
similar phase diagram for a larger (more TF-like) condensate.

\subsection{Stable dynamics}
\begin{figure}[t]
\centering
\includegraphics[width=0.80\columnwidth,clip=true]{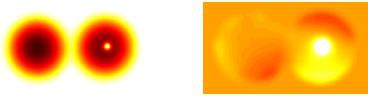}
\caption{(Left) Density $n(x,y)$ for the single vortex scenario with $y_v=1\xi$ and $V_0=0.9\mu$ at a time of 1000 $\xi/c$.  Black (white) corresponds to peak (zero) density.   (Right)  Renormalised density $n(x,y)-n_{\rm VF}(x,y)$ for the same data as above.  The colorscale is $\pm 10\%n_0$.  Each box is of size $64\xi\times32\xi$. }
\label{fig:single_vortex_stable}
\end{figure}

If dissipationless, we would expect the vortex to always remain in
the system.  In Fig.~\ref{fig:phase_diagram_single_vortex} we see
regimes where this is true (dots) but also where the vortex is
unstable and ultimately leaves the system (crosses).  Stable vortex
motion is promoted for large $V_0$, since the well then behaves
likes an isolated harmonic trap, and for low vortex radii, since the
vortex does not feel a strong effect from the far trap. Under this
stable motion the vortex dynamics is akin to that in a single
harmonic trap \cite{Parker2004}: it precesses around the well centre
with approximately constant radius and generates a collective motion
of the background condensate of low amplitude ($\sim5\% n_0$).  In
Fig. \ref{fig:single_vortex_stable} we show a typical snapshot of
the condensate.  The density distribution consists of two
weakly-connected circular condensates and the vortex appears as a
hole (white spot) in the right-hand well.  By subtracting the
time-independent vortex-free density $n_{\rm VF}$ from this density
profile, the collective excitation become clearly visible (Fig.
\ref{fig:single_vortex_stable}).  The precession frequency of the
vortex is $\sim0.2\omega$ for small displacements, and increases
with $r_v$, in good agreement with analytic predictions for vortex
precession in a single harmonic trap in the TF regime
\cite{Svidzinsky2000}.

\subsection{Unstable dynamics}
In cases where the vortex eventually decays from the system (crosses
in Fig.~\ref{fig:phase_diagram_single_vortex}) its initial dynamics
falls into one of two cases. In case I, the dynamics is
characterised by the vortex {\em  crossing over} into the adjacent
well, whereas in case II it is characterised by {\em inducing} a
mirror vortex in the adjacent well.  These effects most commonly
become manifested in the first precession of the vortex in the trap,
but in a minority of cases they may arise after several precessions.

\subsubsection{Case I: vortex crossover}
\begin{figure}[b]
\centering
\includegraphics[width=0.825\columnwidth,clip=true]{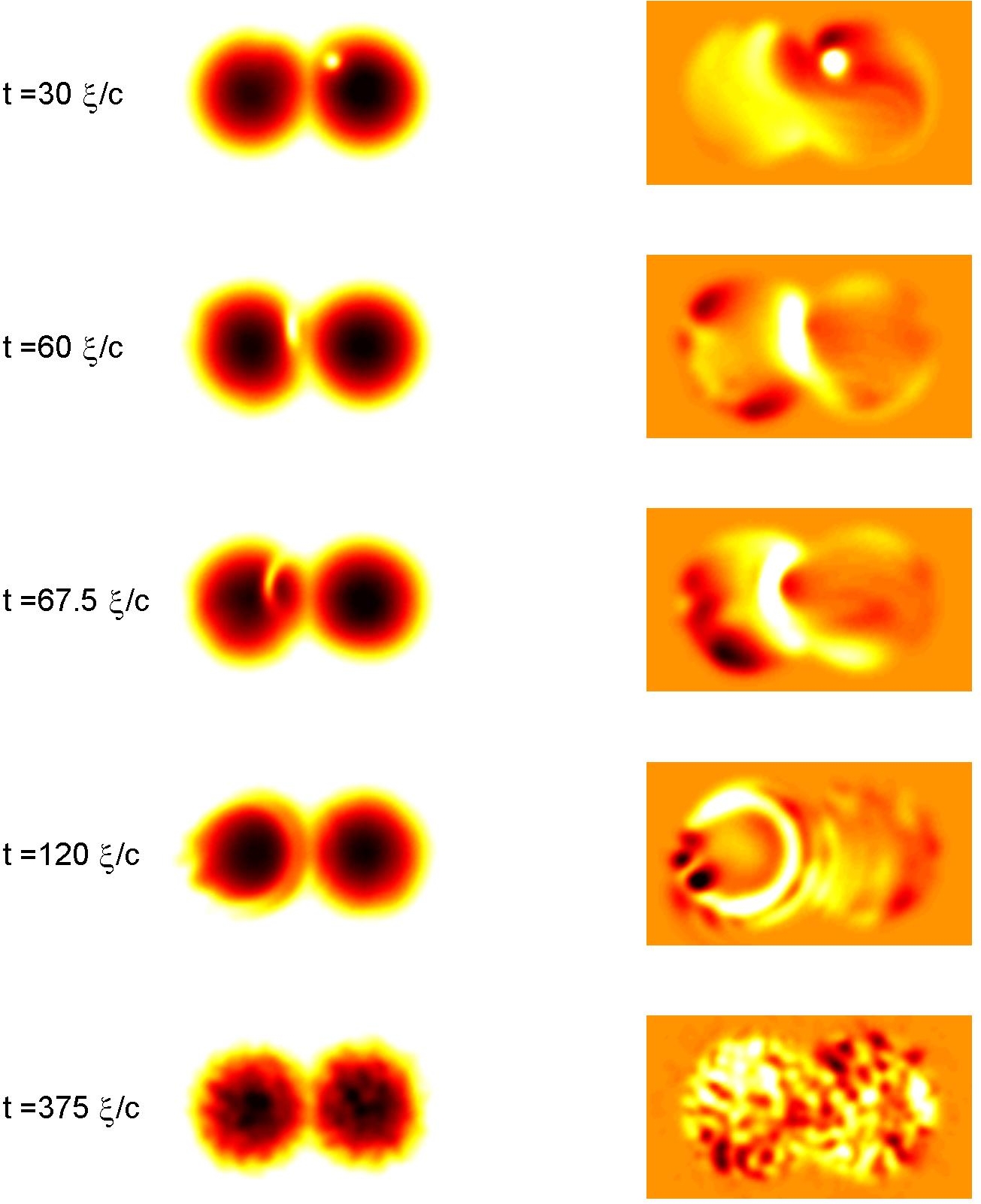}
\caption{
Crossover regime dynamics:
Snapshots of (left) density $n(x,y)$ and (right) renormalized density $n(x,y)-n_{\rm VF}(x,y)$ within the crossover regime, for $r_v=8\xi$ and $V_0=0.6\mu$, at various times. (Spatial and color scales are the same as in Fig.~\ref{fig:single_vortex_stable}).}
\label{fig:crossover_snapshots}
\end{figure}
Case I (vortex crossover) arises in the lightly shaded region in
Fig.~\ref{fig:phase_diagram_single_vortex}.  The vortex can be
expected to travel between the wells when the local potential of the
vortex (the value of the potential at the vortex core) exceeds the
inter-well barrier.  This will tend to occur for large vortex
off-sets and a weak inter-trap barrier.  Put quantitatively, one
would expect cross-over to occur when the vortex density depth $n_v$
is less than or equal to the minimum density at the barrier $n_{\rm
min}$ (one can picture this as when the vortex can just squeeze
through the barrier).  If dissipationless, the density depth of the
vortex will retain its initial value, which we can approximate via
the TF prediction $n_v=n_0(1-r_v^2/R^2)$.  This is a robust
approximation provided that the vortex position is away from the
edge and the barrier, for which the TF density agrees with the
actual density profile to within $0.01n_0$.  This gives the criteria
$r_v/R\geq \sqrt{1-n_{\rm min}/n_0}$ for cross-over to be possible.
We can first approximate $n_{\rm min}$ via its TF prediction $n^{\rm
TF}_{\rm min}=n_0(1-V_0/\mu)$, giving the solid  line.  However, the
onset of cross-over dynamics occurs at considerably lower $r_v$.  We
can expect some deviation to arise from the inaccuracy of using the
TF approximation for $n_{\rm min}$: the TF approximation
underestimates the density at the point of the barrier, as evident
in Fig.~\ref{fig:trap}.

If we instead use the actual value of the ground-state density at
the barrier, we obtain the threshold shown by the dashed line in
Fig.~\ref{fig:phase_diagram_single_vortex}.  This lowers the
prediction for $r_v$, but only slightly and the prediction still
remains considerably greater than the observed threshold.  This
anomaly is likely to arise from the radiation sound from the vortex.
This will have two implications for promoting cross-over dynamics.
Firstly the dissipating vortex will move to lower densities/greater
radial position.  Secondly, the emitted sound generates collective
modes of the BEC which will cause the density at the barrier to
become time-dependent and may promote vortex cross-over during
``high tide''.

In this crossover regime, the ultimate fate of the vortex is to
decay.  In Fig.~\ref{fig:crossover_snapshots} we present an example.
The precessing vortex approaches the barrier ($t=30 (\xi/c)$) and
upon traversing it ($t=60 (\xi/c)$), decays into a high-amplitude
curved pulse of sound ($t=67.5(\xi/c)$).  The sound pulse reflects
off the far left-side of the trap and propagates back through the
trap ($t=120 (\xi/c)$).  Following many reflections and diffractions
in the trap, the sound pulse becomes randomised, ultimately forming
an isotropic sound field ($t=375(\xi/c)$).

In other cases, the vortex undergoes a more gradual decay, passing
many times between the wells.  NB more generally, we can observe
some cases where the vortex undergoes stable cross-over dynamics, as
we will see later in Section VII.

\subsubsection{Case II: vortex induction}

\begin{figure}[h]
\centering
\includegraphics[width=0.85\columnwidth,clip=true]{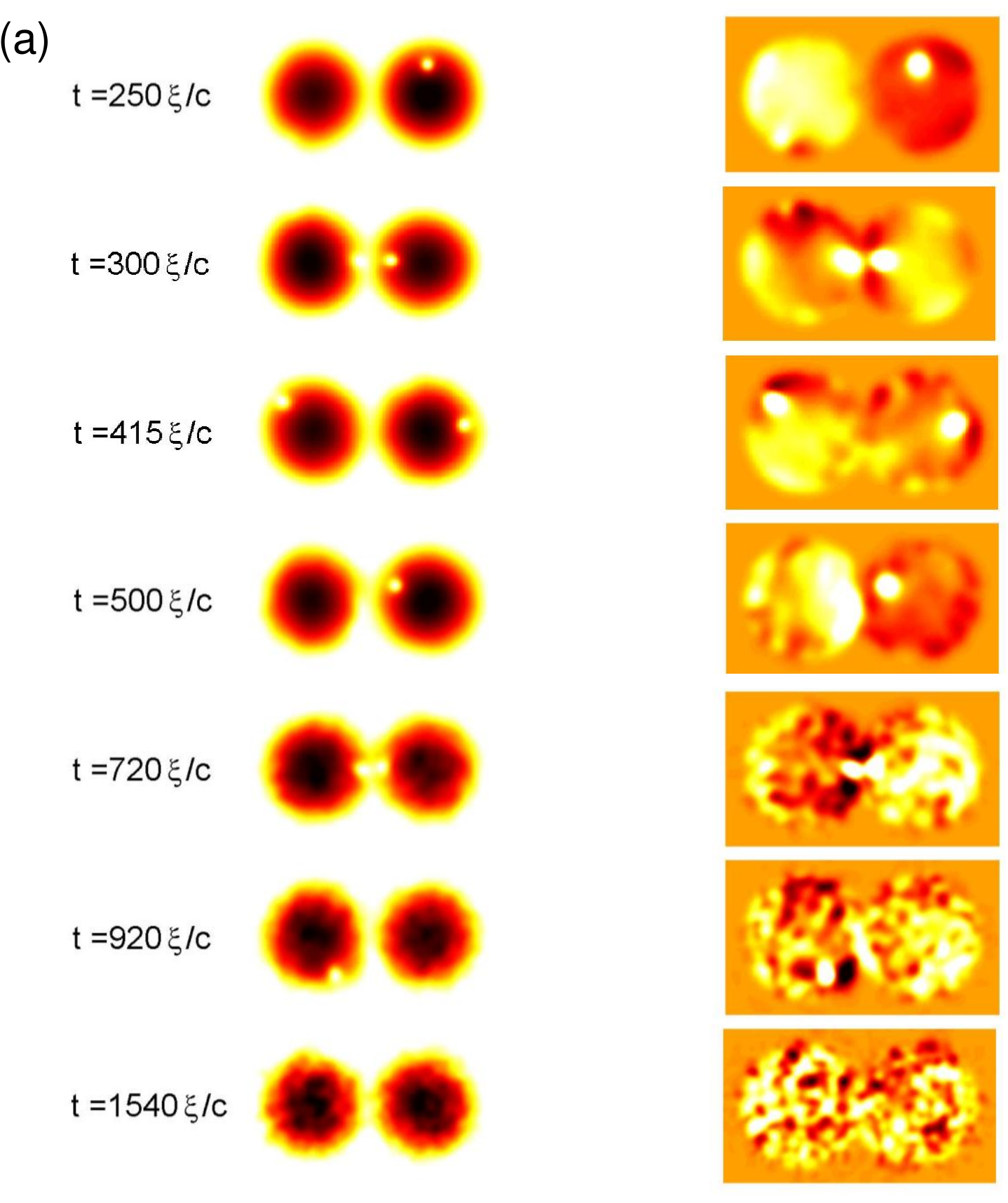}\\
\includegraphics[width=0.825\columnwidth,clip=true]{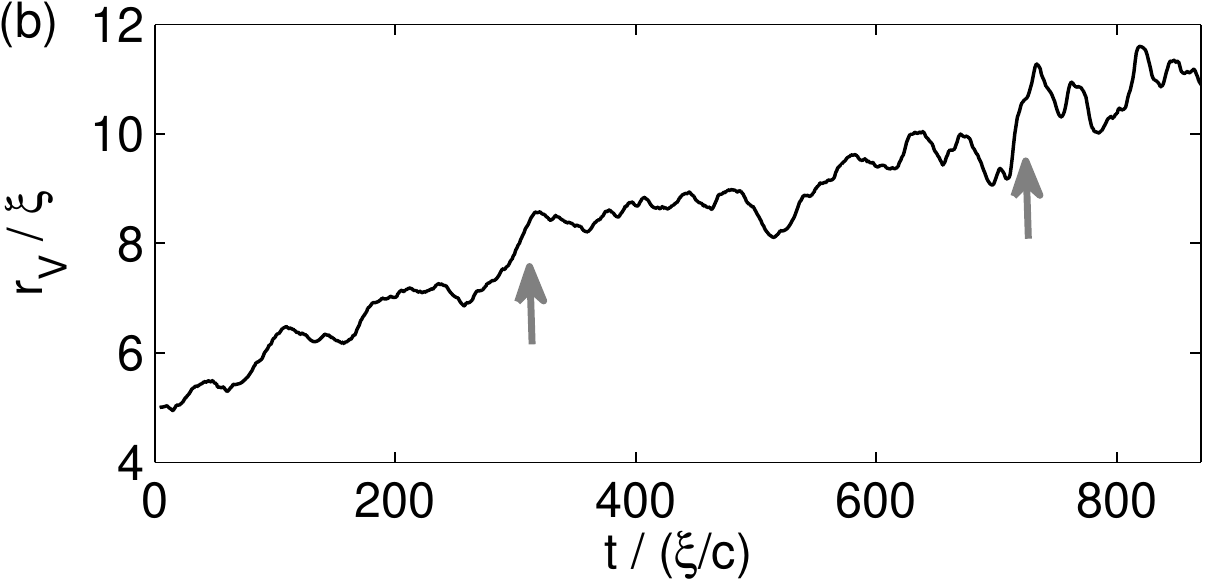}
\caption{Inductive regime dynamics:
 (a) Density $n(x,y)$ (left column) and renormalised density $n(x,y)-n_{\rm VF}(x,y)$ (right column) at various times. Parameters: $r_v=5\xi$ and $V_0=0.9\mu$ (box size and colorscale are the same as in Fig.~\ref{fig:single_vortex_stable}). (b) Evolution of the radial position of the vortex $r_v$; arrows indicate the points at which vortex induction takes place. }
\label{fig:single_vortex_induction}
\end{figure}

The second case of unstable dynamics (case II: vortex induction),
characterised by the initial vortex inducing a second vortex in the
far-well, arises in the dark shaded region in Fig.
\ref{fig:phase_diagram_single_vortex}.  Snapshots of a typical
evolution are shown in Fig.~\ref{fig:single_vortex_induction}.  As
the vortex precesses close to the adjoining well ($t=300 \xi/c$), a
mirror vortex becomes excited on the opposite side of the inter-well
barrier.  This vortex has the opposite charge to the original vortex
and precesses in the opposite direction around its trap ($t=415
\xi/c$).  This early induction is the characteristic of this regime
of dynamics.  The subsequent dynamics can vary widely. In the
example, the initially-induced vortex disappears ($t=500 \xi/c$), a
second vortex is induced ($t=720\xi/c$) and then the original
right-hand vortex disappears ($t=920 \xi/c$).  In other cases the
original vortex may disappear as it creates its mirror vortex, while
sometimes the induced vortex may itself induce a vortex in the
right-hand well.  As a result there can arise periods of time where
multiple vortices can appear in the system.  For all cases we
observe the growth of a tempestuous sound field during the vortex
motion, set up by the sound emission from the accelerating vortices.
All of the vortices ultimately dissipate and disappear into an
energetic and isotropic sound field
(Fig.~\ref{fig:single_vortex_induction}(b) at $t=1540 \xi/c$).

Up until the point when it disappears, the original vortex drifts
outwards, as shown in Fig.~\ref{fig:single_vortex_induction}(b).
Most of the noise in $r_{ v}(t)$ is due to the buffetting effect
that the sound field has on the vortex.  However, the sizeable jumps
in $r_v$ at $t\sim300$ and $700~(\xi/c)$ (highlighted by arrows)
occur as a vortex is induced in the far well, indicating the
transfer of energy from the original vortex to create the new one.

This region of the parameter space occurs for a range of barrier
heights in the vicinity of $V_0=\mu$, for which the condensate
channel is of low density.  The energy to create a vortex depends on
the local density.  As the barrier height is reduced below $\mu$,
the density in the barrier region increases and we have confirmed
with numerical simulations that the energy cost to create a vortex
here increases sharply.

\section{Cross-talk of two vortices}
\begin{figure}[b]
\centering
\includegraphics[width=1\columnwidth,clip=true]{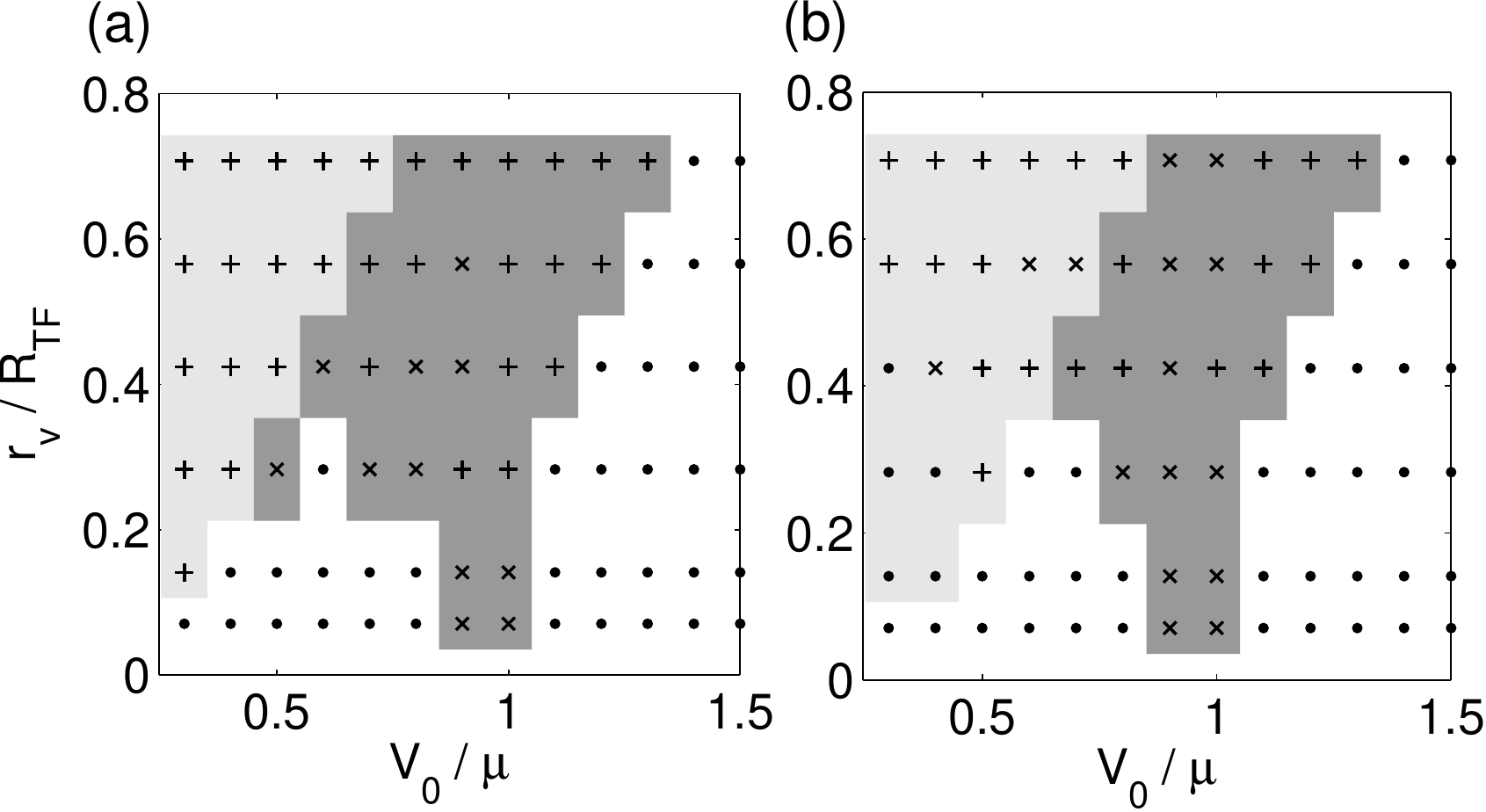}
\caption{Phase diagram in $r_{v}-V_0$ of the final state (after $5000 (\xi/c)$) of the two vortex system for (a) same polarity and (b) opposite polarity of the vortices.  The final state is either two stable vortices (dots), one stable vortex (pluses) or no vortices (crosses).  The second vortex is initially positioned at the centre of the left-hand well.  Light and dark shading indicates case I (crossover) and case II (induction) of the unstable dynamics.}
\label{fig:two_vortex_phase_diagram}
\end{figure}
\begin{figure*}
\centering
\includegraphics[width=1.4\columnwidth,clip=true]{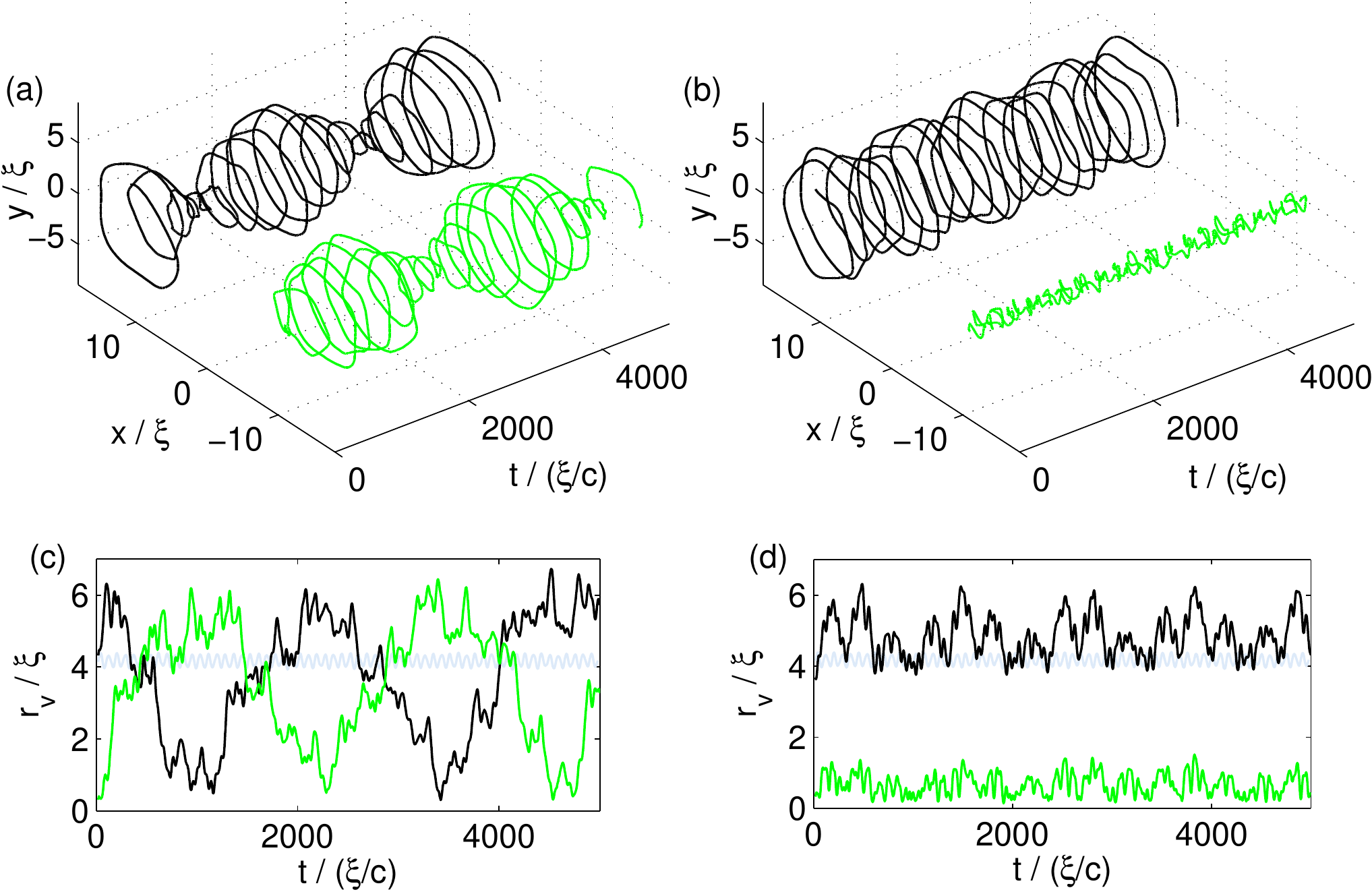}
\caption{Vortex dynamics in the two-vortex system with
(a) the same polarity and (b) different polarity.
Parameters: $r_1(t=0)=4\xi$, $V_0=0.6\mu$.  Black (green/dark grey) lines correspond to vortex 1 (2).  (c) and (d) plot the corresponding vortex radii, and additionally include the evolution of a single vortex in an isolated harmonic trap (light blue/grey line) with initial radius $r_{\rm v}(t=0)=4\xi$.}
\label{fig:double_vortex_radii}
\end{figure*}
We now extend to the case where there is initially a vortex in {\it
each} well to examine the possibility of sound-induced interaction
between vortices. We continue to employ the idealized double
harmonic trap geometry; we will demonstrate the same phenomena in a
realizable double trap geometry in Section \ref{sec:exp}.   The
additional vortex (denoted ``vortex 2") is created at the centre of
the left-hand trap.  Since it feels the velocity field of vortex 1
it will is not perfectly stationary.  The displacement of vortex 1,
denoted $r_{v}$, and the height of the inter-well barrier, $V_0$,
are again key to the dynamics.  The latter parameter now
additionally controls the transfer of sound between the wells, since
sound waves have an energy of around $\mu$.  We can also expect
sensitivity to the relative polarity of the vortices.  Figure
\ref{fig:two_vortex_phase_diagram} shows the phase diagram of the
system for (a) vortices of the same polarity and (b) vortices of
opposite polarity.

The phase diagrams are similar to the single vortex case
(Fig.~\ref{fig:phase_diagram_single_vortex}) and the vortex polarity
only has a small effect on the final state of the system. There are
two stable regions, one for weak barriers and low displacements, and
the other for large barriers.  We also see a cross-over regime for
weak barriers and large vortex displacements in which vortex 1 tends
to cross into the other well, and an induction regime for barrier
heights centered around $\mu$ in which vortex 1 induces another
vortex in the opposite well.  Where vortex instability does occur,
the end state is either the persistence of a single vortex or no
vortices at all.  We cannot make any general comments about what
favours these two end states: the dynamics that can develop are
sufficiently complex that the end state is not readily
deterministic.

Note that in Fig.~\ref{fig:two_vortex_phase_diagram}(b) there
exists stable dynamics in the cross-over regime.  In this case
vortex 1 perpetually traverses both wells in a stable manner while
vortex 2 remains localised in the centre of its well.

In order to investigate the cross-talk between vortices, that is
their sound-mediated interactions, we focus on the stable regimes
and in particular the stable area occurring for low barrier heights,
for which one can expect unimpeded motion of sound between the
wells. Figure \ref{fig:double_vortex_radii} present the vortex
dynamics for $r_{v}=4\xi$ and $V_0=0.6\mu$.  When the vortices have
the same charge (Fig.~\ref{fig:double_vortex_radii}(a)) the vortices
periodically spiral inwards and outwards, and remain out of phase
with each other.  The change in vortex radial position
(Fig.~\ref{fig:double_vortex_radii}(c)) is large (the vortices
oscillate between the trap centre and a radius of around $6 \xi=0.43
R_{\rm TF}$), demonstrating a significant transfer of energy between
the vortices.  The energy exchange occurs over a timescale of
$\sim2000 (\xi/c)$, which is around 8 precessions of the vortex in
the trap (the precession period is $\sim 270 (\xi/c)$).  In the
analogous situation for the dark soliton, the energy transfer is
much slower, occurring over many tens of soliton oscillations
\cite{Allen2011}.  The periodic change in vortex radius
(Fig.~\ref{fig:double_vortex_radii}(c)) is composed of sharp steps,
suggesting that the energy emission and absorption of the vortices
occurs in packets rather than continuous.

In Fig.~\ref{fig:double_vortex_radii}(c) we also present, for
comparison, the evolution of a single vortex in an isolated harmonic
trap (light blue/grey line).  The vortex maintains an approximately
constant radius, with only small-scale modulations due to its
re-interaction with its emitted sound and collective modes
\cite{Parker2004}.

When the vortices have opposing polarity
(Fig.~\ref{fig:double_vortex_radii}(b) and (d)), no significant
transfer of energy is observed and the vortices precess with
approximately constant radius. The vortex motion does undergo
modulations but these are not out-of-phase and so do not constitute
a direct exchange of energy. Rather the modulations arise from the
back-action of the randomised sound field on the vortices.  These
modulations are larger than the corresponding ones experienced by a
single vortex in a harmonic trap, indicative of the greater sound
density in the double well system.

The significant transfer of energy between like-charged vortices and
insignificant transfer between unlike-charged vortices is consistent
across the stable region of the parameter space at low barrier
heights.  In the region of stable dynamics at high $V_0$ we do not
observe a significant transfer in energy between the vortices since
the transfer of sound across the barrier becomes prohibited.

It would appear that the vortices periodically drive energy into
each other via their emitted sound.  However, we must rule out that
this effect arises from the long-range interaction between vortices
due to their superimposed velocity fields.  To this aim, we will
next attempt to drive a vortex via sound waves generated by
artificial means - a moving localised barrier.

\section{Precessing obstacle}
We now replace the left-hand vortex with a precessing obstacle.
Note that we continue to employ the idealized double harmonic trap;
the same effects will be demonstrated in an experimentally
realizable trap in Section \ref{sec:exp}.  The obstacle corresponds
to a time-dependent potential,
\begin{eqnarray}
V_{\rm ob}(x,y,t)&=&A_{\rm ob}
\\ \times &\exp&\left[{-\frac{\left\{x+x_c-x_{\rm ob}(t)\right\}^2+\left\{y-y_{\rm ob}(t)\right\}^2}{\sigma^2}}\right] \nonumber,
\end{eqnarray}
where,
\begin{eqnarray*}
x_{\rm ob}(t)=\pm r_{\rm ob} \sin(\omega_{\rm ob}t+\phi_{\rm ob})\\
y_{\rm ob}(t)=r_{\rm ob} \cos(\omega_{\rm ob}t+\phi_{\rm ob}).
\end{eqnarray*}
This represents a Gaussian-shaped barrier, of amplitude $A_{\rm ob}$
and width $\sigma$, moving in a circular path around the centre of
the left-hand well with radius $r_{\rm ob}=x_{\rm ob}^2+y_{\rm
ob}^2$.  Such an obstacle can be induced experimentally via a
blue-detuned laser beam \cite{Neely2010,Raman2001}. For the ``-"
(``+") case above, the barrier moves in the same (opposite)
direction as the vortex. $\phi_{\rm ob}$ represents the relative
phase between the initial position of the obstacle and the vortex
(e.g. for $\phi_{\rm ob}=0$ the obstacle and vortex start at
mirror-opposite positions in their wells).

We attempt to generate sound that mimics the sound generated by a
vortex and will thus employ precession frequencies similar to those
of a vortex ($\sim 0.23\omega$ for our system). This is sufficiently
low to prevent the obstacle exceeding the superfluid critical
velocity and nucleating vortices
\cite{Jackson2000,Onofrio2000,Raman2001,Neely2010}. We have
confirmed via numerical simulations of the precessing obstacle (with
no vortices imposed) that its only effect is to generates spiral
sound waves with typical amplitude $\sim1\% n_0$.

\begin{figure}[t]
\centering
\includegraphics[width=0.725\columnwidth,clip=true]{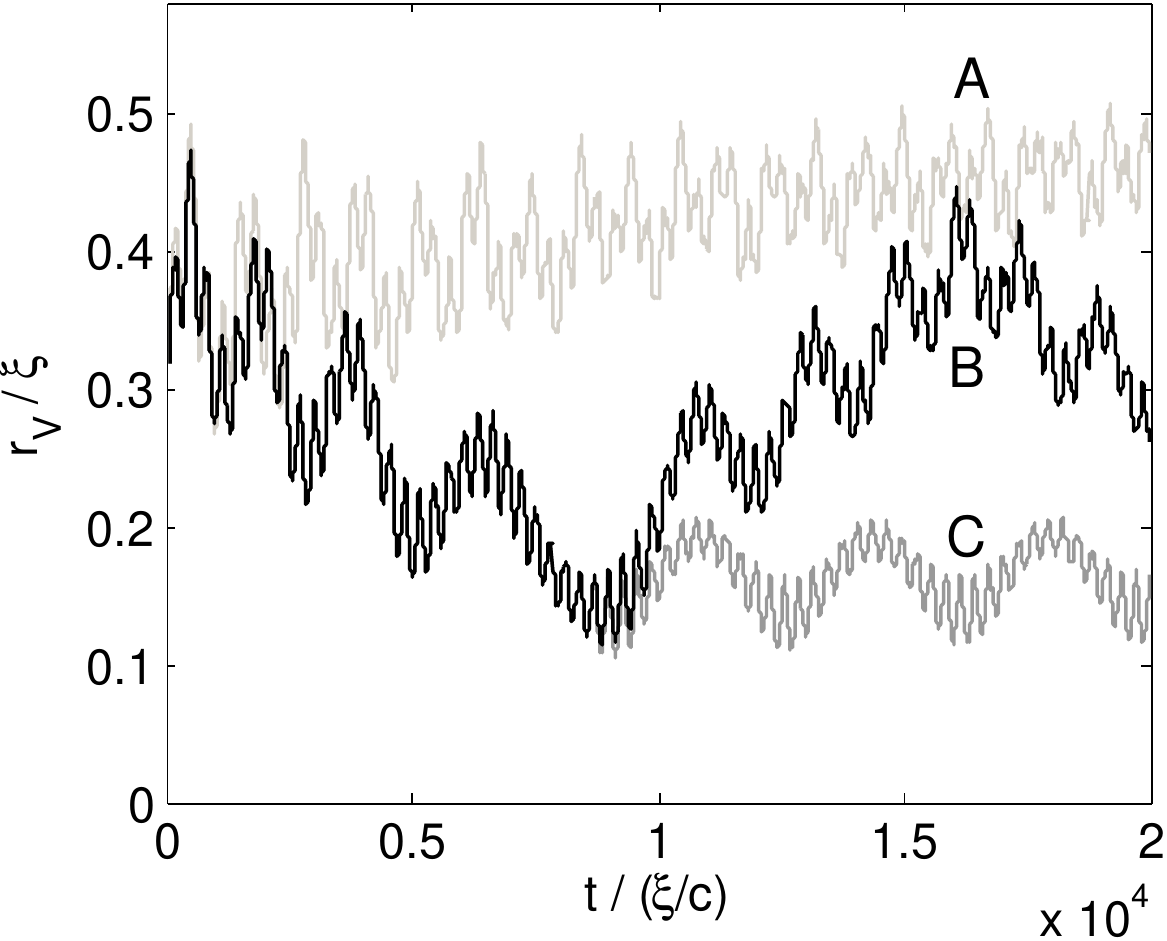}
\caption{Evolution of the vortex radius under the influence of a driving obstacle.   Black line (B): the obstacle moves in the same direction as the vortex.  Light grey line (A): the obstacle moves  in the opposite direction to the vortex.  Dark grey line (C): the barrier moves  in the same direction as the vortex but its motion is terminated at t=$8800 (\xi/c)$.  Parameters: $r_1(t=0)=4\xi$, $\sigma=2\xi$, $r_{\rm ob}=4\xi$, $\phi_{\rm ob}=0$ and $\omega_{\rm ob}=0.234 \omega$ and $V_0=0.6\mu$. }
\label{fig:moving_obstacle_example}
\end{figure}

We mimic the system parameters employed in
Fig.~\ref{fig:double_vortex_radii} [$r_v=4\xi$ and $V_0=0.6\mu$] and
additionally employ an obstacle potential defined by width
$\sigma=2\xi$, amplitude $A_{\rm ob}=\mu$,  phase $\phi_{\rm ob}=0$,
path radius $r_{\rm ob}=4\xi$ and frequency $\omega_{\rm ob}=0.234
\omega$ (these parameters are chosen as they optimise the energy
transfer, as shown below).   First consider the case when the
barrier moves in the same direction as the vortex (solid black line
in Fig.~\ref{fig:moving_obstacle_example}).  We clearly see that the
vortex radial position oscillates in time, i.e., there is periodic
driving of energy into the vortex.  This is analogous to the
predicted parametric driving of a matter-wave dark soliton
\cite{Proukakis2004}.  The energy transfer due to the obstacle is
more gradual than that induced by another vortex (i.e., compare to
Fig.~\ref{fig:double_vortex_radii}), with the exchange occurring
over a time of $\sim12000$ ($\xi/c$) or $\sim 40$ precessions.
Conversely, if the barrier moves in the opposite direction to the
vortex (solid green/grey line) no driving is observed (although a
small outward shift of the vortex suggests a small loss of energy in
this case).  This is qualitatively the same behaviour as we observed
earlier when the obstacle in replaced by another vortex.  However,
the timescale of the energy transfer is considerably slower than
with the obstacle.  This shows that the vortex is a much more
efficient source of sound to drive another vortex.

Returning to the case where the obstacle moves in the same direction
as the vortex, we find that if the obstacle motion is terminated
(becoming a stationary obstacle) when the vortex is at its minimum
radial position (dotted black line in
Fig.~\ref{fig:moving_obstacle_example}), it maintains this minimal
radial position (subject to some oscillations about this radius). In
this manner we can parametrically drive net energy into the vortex.
Indeed, if we terminate the obstacle motion at any point the vortex
approximately maintains that radial position.

The minimum radius achieved $r_{\rm min}$ provides a measure of the maximal amount of energy driven into the vortex.  In Fig.~\ref{fig:moving_obstacle_resonance} we show how this quantity varies as a function of the obstacle parameters $\omega_{\rm ob}, $$V_{\rm ob}$, $r_{\rm ob}$, $\sigma$ and $\phi_{\rm ob}$.  For each case we observe a clear resonance that minimises $r_{\rm min}$ and maximises the energy that can be driven into the vortex:

\begin{figure}[t]
\centering
\includegraphics[width=0.9\columnwidth,clip=true]{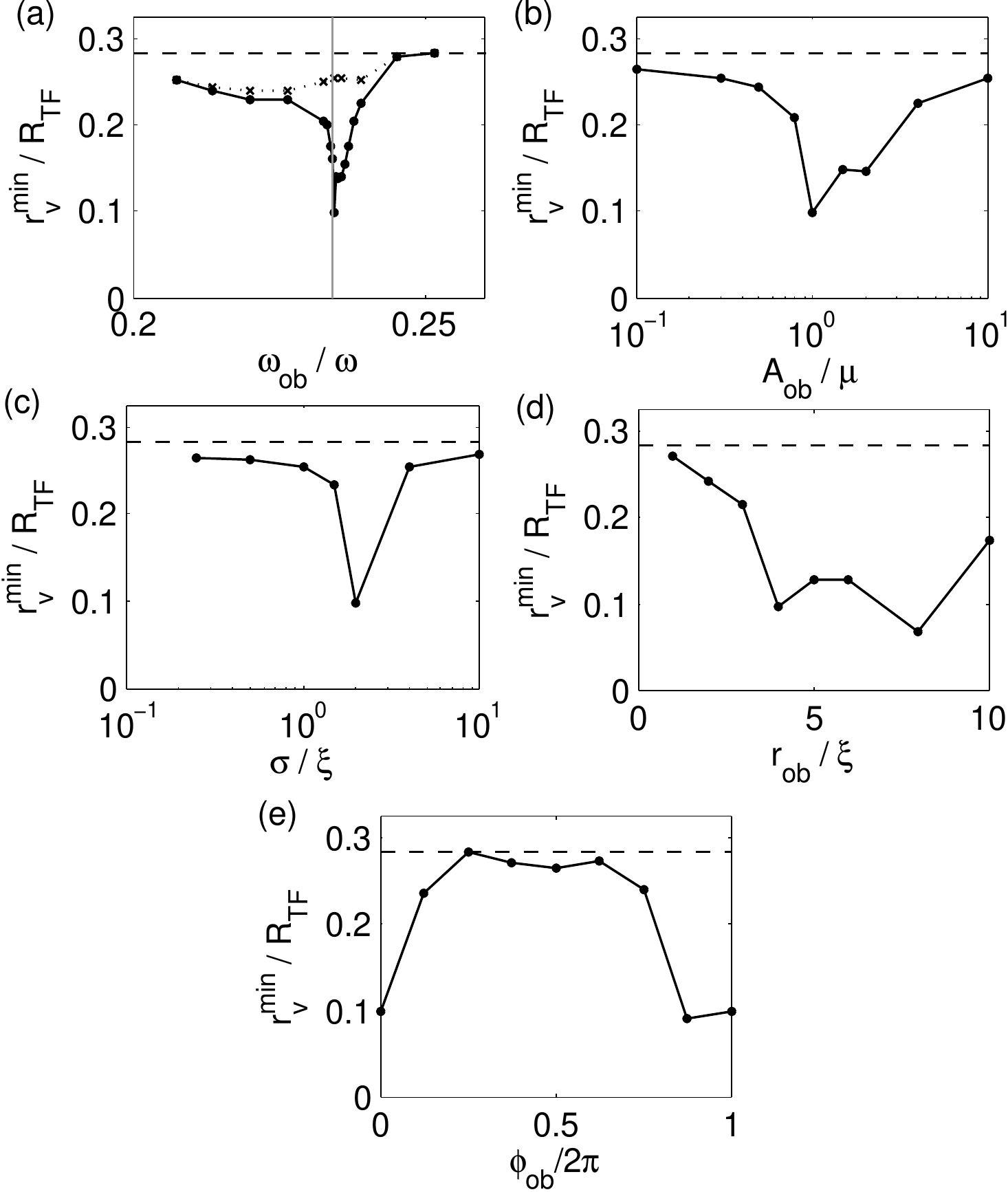}
\caption{Minimum radial position achieved by the vortex under the influence of a precessing Gaussian barrier during a total simulation time of $2\times10^4 (\xi/c)$.  The vortex is initially at $r_v=4\xi$ and the inter-well barrier height is $V_0=0.6\mu$.   The obstacle parameters are based around the values $\omega_{\rm ob}=0.234\omega$, $A_{\rm ob}=\mu$, $\sigma=2\xi$, $r_{\rm ob}=4\xi$ and $\phi_{\rm ob}$, and each is varied in turn from plot (a)-(e).   In (a), we present results for co-rotating (solid line with dots) and anti-rotating (dashed line with crosses) system. In (b)-(e) we consider only the co-rotating system.  In each plot the initial vortex radial position is shown (dashed horizontal line).  In (a) we show the precession frequency of the undriven vortex (grey vertical line).  }
\label{fig:moving_obstacle_resonance}
\end{figure}

\begin{itemize}
\item{$\omega_{\rm ob}$ [Fig.~\ref{fig:moving_obstacle_resonance}(a)]: When the obstacle direction is the same as the vortex (black line/dots) we see a clear frequency resonance, with maximal energy being driven into the vortex for a precession frequency close to that of the undriven vortex (vertical dotted line).  For the opposite case, the minimum radius achieved undergoes no significant resonance.  For this reason, we subsequently only consider the co-rotating case.}

\item{$V_{\rm ob}$ [Fig.~\ref{fig:moving_obstacle_resonance}(b)]: $r_{\rm min}$ is minimised for an obstacle height of $\sim \mu$.}

\item{$\sigma$ [Fig.~\ref{fig:moving_obstacle_resonance}(c)]: A barrier width $\sigma\sim2\xi$ best promotes energy driving.}

\item{$r_{\rm ob}$ [Fig.~\ref{fig:moving_obstacle_resonance}(d)]: The barrier position shows a double resonance at $4\xi$ and $8\xi$.}

\item{$\phi_{\rm ob}$ [Fig.~\ref{fig:moving_obstacle_resonance}(e)]: Maximal energy driving occurs when the vortex and obstacle begin in phase or with the obstacle slightly lagging the vortex.}
\end{itemize}

\section{Discussion}
\subsection{Cross-over and induction dynamics}
Our precursive study of a single vortex in a double trap revealed
cross-over and inductive dynamics.  The generation of a vortex state
in a neighbouring well of a double-well trap has been studied in
\cite{Salgueiro2009}. However, there the vortex-containing BEC
partially tunnelled into the adjacent empty well, and so is a
distinct creation mechanism to the one observed here.  While these
regimes are interesting in their own right, from the perspective of
exploring cross-talk and parametric driving of vortex they pose
limitations on the parameter space in which vortices are stable.

\subsection{Cross-talk between vortices}
Our main results demonstrate cross-talking between two vortices.
When the vortices have the same polarity and different initial
positions, they undergo periodic exchanges of energy with each
other, transferred via sound waves.  We may view this as the driving
of a vortex by sound from another vortex.  The transfer of energy is
rapid: a full energy cycle typically occurs within $10$ vortex
precessions.

Sound is emitted from each vortex due to its acceleration in the
trap. This emitted sound has a quadrupolar radiation pattern and
forms an outwardly propagating spiral wave \cite{Parker2004a} that
carries angular momentum of the same orientation as the originating
vortex (to conserve angular momentum).  Upon passing into the
opposing well it imparts some of its energy and angular momentum to
the opposing vortex.  If the angular momentum carried by the sound
waves is of the same orientation as the receiving vortex (which is
to say that the vortices are of the same polarity), it serves to
increase the angular momentum and energy of this vortex, causing it
to move towards the high density trap centre. By the same simple
argument, one would expect that when the vortices are of opposite
polarity, the energy and angular momentum of the opposing vortex
would be reduced. However, no exchange of energy is observed when
the vortices have opposing polarity.  We will return to this anomaly
below. The cross-talk between vortices is analogous to that
occurring for dark solitons in the 1D analog of this system
\cite{Allen2011}. However, where the energy transfer for solitons
occurs over many tens of oscillations, the energy transfer for
vortices can occur in less than 10 oscillations.  Thus the vortex
system appears a beneficial platform to experimentally observe and
explore these related sound phenomena.

\subsection{Parametric driving of a vortex}
By moving a localised Gaussian obstacle through one well, we are
able to generate angular-momentum carrying sound waves.  When this
angular momentum is of the same sign as the vortex, this acts to
drive energy into the vortex, but when it of opposite sign it is
essentially invisible to the vortex, as seen above for the
vortex-vortex transfer.  Under constant driving, the vortex energy
oscillates periodically and the oscillation amplitude can be
resonantly tuned via the obstacle parameters.   The driving is most
effective when the time-dependent density perturbation created by
the obstacle closely matches that of the vortex, i.e. the same
width, precession frequency and amplitude. However, the
vortex-obstacle energy transfer is considerably slower than
vortex-vortex transfer, suggesting that the most ``natural" sound to
drive a vortex is that from another vortex.

An analogous set-up was studied in \cite{Caradoc-Davies1999}, except
the vortex and obstacle resided in the same trap.  There the
precessing obstacle periodically formed a vortex at the condensate
edge and drove it into the centre.  The dynamics were interpreted as
nonlinear Rabi cycling from the ground state to the first vortex
state. Similarly, we may interpret our observations as oscillations
between a high and low energy vortex state.  Our results suggest
that sound waves may have played a central role in transferring
energy from the obstacle to the vortex in \cite{Caradoc-Davies1999}.

Our findings mirror those relating to driving of dark matter-wave
solitons in \cite{Proukakis2004}. There, 1D oscillating paddles were
employed to impart linear momentum to the condensate sound field
which in turn led to energy being driven into the dark soliton.
Resonant driving occurred when the paddles were oscillated at close
to the frequency of the undriven soliton motion. Furthermore, when
the drive was switched off the soliton maintained that energy (in
the absence of phenomenological dissipation), subject to
oscillations from the background density excitations.

\subsection{Insight into sound absorption in trapped and homogeneous systems}
Using an acoustic ray model, Nazarenko {\it et al.}
\cite{Nazarenko1994,Nazarenko1995} find that certain trajectories of
sound waves incident on a vortex line can spiral into the vortex
core, imparting their energy to the vortical flow.  Despite this and
our observations herein,  sound absorption is not cited to play a
significant role in superfluid Helium systems, e.g. quantum
turbulence \cite{Barenghi2001,Vinen2010}.

Consider an infinite homogeneous system containing a tangle of
vortex lines.  Within this, consider an element of vortex line
emitting a pulse of sound due to some acceleration. From the point
of emission, the sound spreads out radially.  This rapidly dilutes
its energy and momentum density, and thus its capacity to influence
a line element in its path.   Superimpose many such events from many
randomly oriented vortex lines and the combined sound field will be
tends towards being isotropic with no net angular momentum.  As such
it is not surprising that sound absorption is insignificant in these
systems.

Now consider our trapped BEC. Sound radiated outwards by an
accelerating vortex will eventually reflects off the trap wall and
become partially focussed towards the trap centre. While the
effectiveness of the focussing will vary with vortex position and
trap shape, it will nonetheless lead to a greater and more sustained
sonic energy and momentum density than in a homogeneous system,
which a suitably placed vortex may be able to gain from. The
focussing effect will, however, be shortlived as the sonic angular
momentum will become randomised after several reflections in the
trap. We see this in Fig.~\ref{fig:crossover_snapshots} where a
large sound pulse with net momentum is rapidly randomised into an
isotropic sound field. This further explains the rapid equilibration
of the vortex energy when the driving is terminated. These simple
arguments suggest that sound absorption may play a significant role
in trapped condensates.

Irrespective of the source of the sound, we observe transfer of
energy between sound and vortex only when they have the same sign of
angular momentum.  We may interpret this in terms of the acoustic
ray picture \cite{Nazarenko1994,Nazarenko1995} which predicts that
trajectories of sound spiral into the vortex core and dump their
energy and momentum.  There the sound was modelled as plane waves
while in the BEC the incident sound is a spiral wave carrying
angular momentum. The crucial aspect is that, when the sound waves
have the same angular momentum as the vortex, they naturally wrap
around the vortex, promoting them to spiral into the core.
Conversely, when the sound has opposite angular momentum, it will
tend to be deflected and repelled by the vortex, unable to impart
its angular momentum.  In essence, the vortex is invisible to sound
waves of opposite angular momentum.

In our quasi-2D system, the vortex line is rectilinear and can only
raise its energy by moving to regions of higher density. In 3D
system, the vortex can additionally increase its energy by
increasing its line length or developing excitations, e.g. Kelvin
waves.  It is not clear which route to absorb energy would be
favoured in 3D, although the orientation of the incoming sound is
likely to play a deciding role.  In systems of multiple vortices,
the position of the vortices provides another route to modify the
total vortical energy.  Furthermore, symmetric excitations such as
vortex rings and vortex-antivortex pairs carry linear momentum and
so can be expected to interact predominantly with sound waves of
linear momenta, much like dark solitons.

\section{Proposed experimental realization}
\label{sec:exp}
\begin{figure}[b]
\centering
\includegraphics[width=0.75\columnwidth,clip=true]{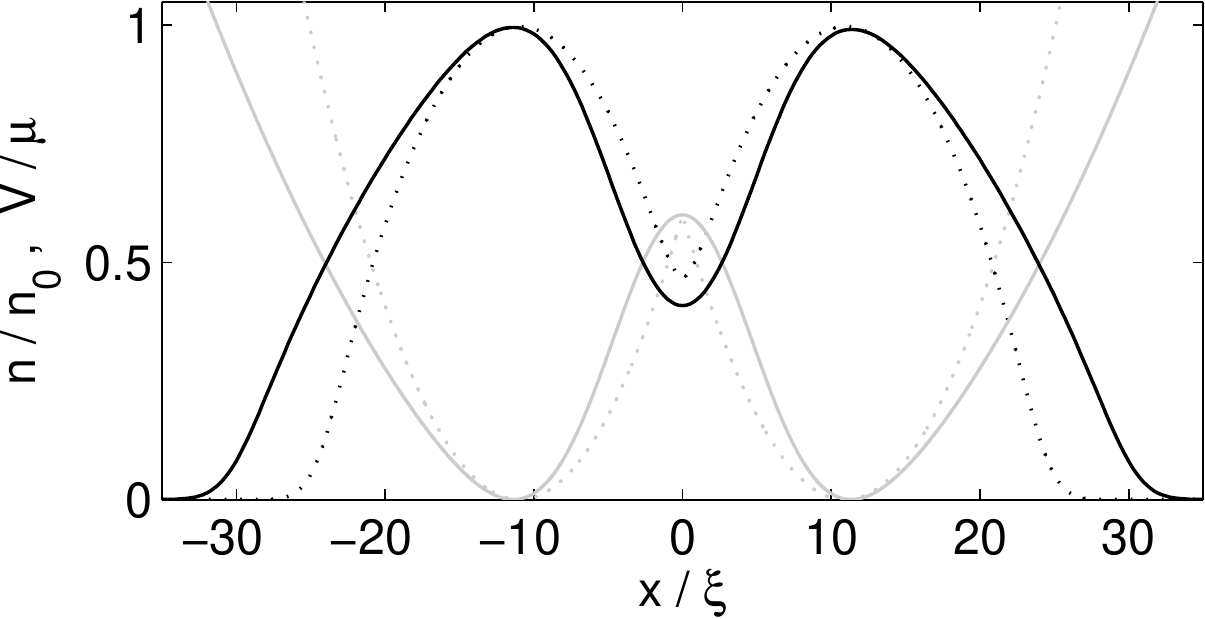}
\caption{The experimentally realizable split-trap (grey solid line)
and its vortex-free density profile (black solid line) along the
$y$-axis.   The analogous potential (grey dashed line) and density
(black dashed line) for the double harmonic trap are shown.}
\label{fig:split_trap}
\end{figure}

\subsection{Set-up details}

The double trap used thus far (defined by Eq.\
(\ref{eqn:double_harmonic_potential})) is idealized and not
experimentally realizable.  Here we will demonstrate that the same
qualitative phenomena occur in experimentally-realizable traps.  We
approximate the idealized double harmonic trap by considering a
single elliptical harmonic trap which is split by a central Gaussian
barrier.  The ``split trap" has the form,
\begin{eqnarray}
V(x,y)=\frac{m}{2}\left[\left(\frac{\omega}{2}\right)^2 x^2+\omega^2 y^2 \right]
+V_{\rm B}e^{-x^2/2d^2}-V_{\rm min},
\end{eqnarray}
where,
\begin{eqnarray}
V_{\rm min}=-m\omega^2 d^2 \left[\ln\frac{V_{\rm B}}{m\omega^2 d^2}+1 \right],
\label{eqn:split_trap}
\end{eqnarray}
is an offset introduced such that the minimum of the potential is zero.  The height of the barrier relative to the trap minimum $V_0$ is related to these parameters via $V_0=V_{\rm B}-V_{\rm min}$.  We set the barrier width to be $d=5\xi$.  Figure~\ref{fig:split_trap} compares the split trap of Eq.\ (\ref{eqn:split_trap}) with the idealized double harmonic trap of Eq.\  (\ref{eqn:double_harmonic_potential}).

Experimentally, the harmonic trap can be formed via magnetic or
optical fields, while the Gaussian barrier is formed by a
blue-detuned laser beam aligned along the $y$-axis.   As previously,
the system we simulate satisfies $\mu=10 \hbar \omega$.
\begin{figure}[b]
\centering
\includegraphics[width=0.925\columnwidth,clip=true]{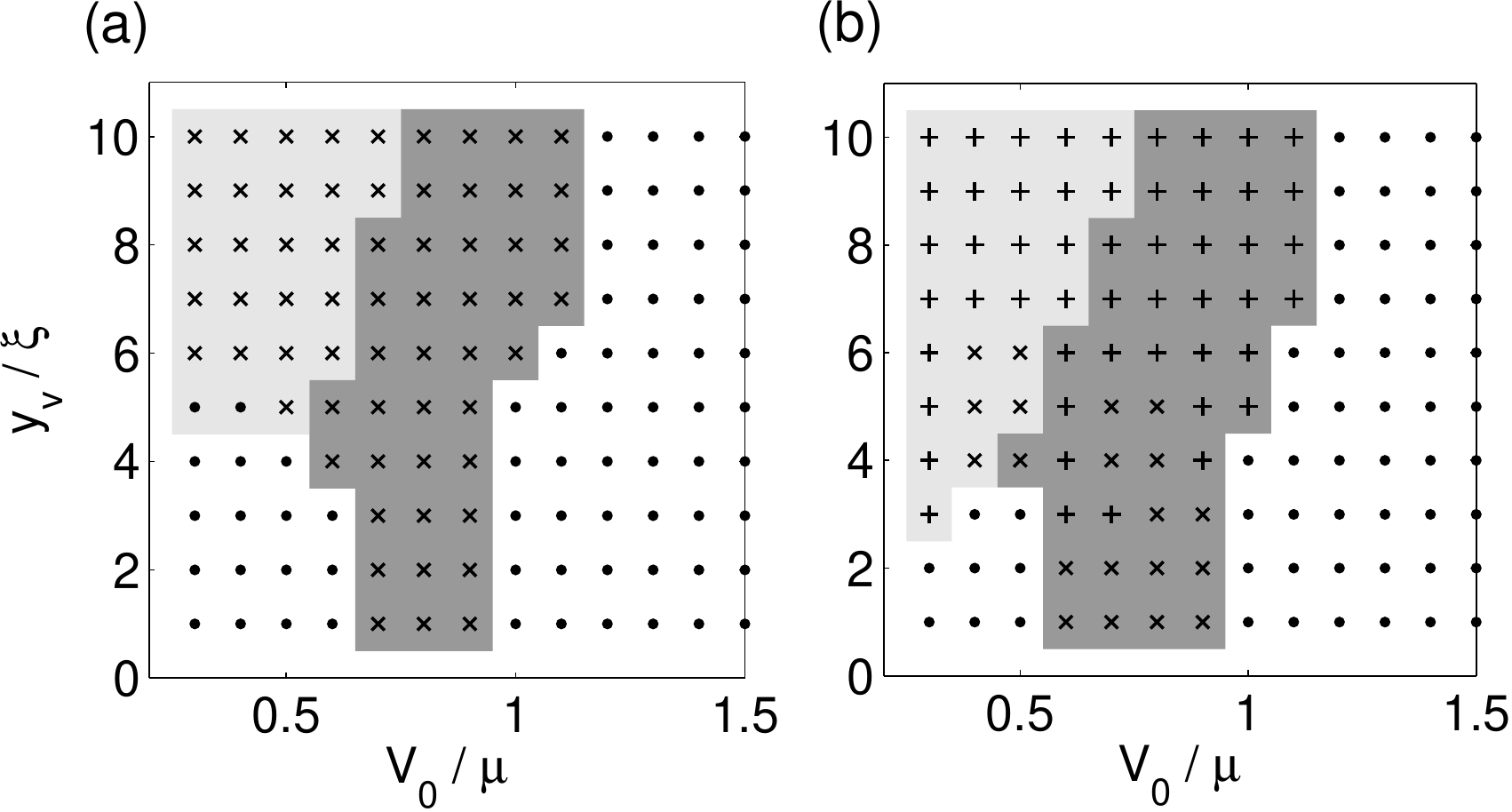}
\caption{Phase diagrams for vortex dynamics in the split trap as a
function of vortex displacement $y_v$ and barrier height $V_0$. (a)
A single vortex in the right-hand well, displaced by $y_{v}$ from
the well centre.  Dots (crosses) correspond to the final state
[after $5000$ $(\xi/c)$] being a vortex (vortex-free) state.  (b) A
vortex in each well, of the same polarity.  The final state is
either two vortices (dots), one vortex (pluses) or no vortices
(crosses).  In (a) and (b) the light-shaded (dark-shaded) region
corresponds to case I (II) of vortex dynamics. }
\label{fig:two_vortex_phase_diagram_split_trap}
\end{figure}
Let us give some typical values for our units.  We will assume some
typical values: a 2D peak density of $n_0=10^{14}$m$^{-2}$, an axial
trap frequency of $\omega_z=2\pi\times1000$ Hz and a radial trap
frequency of $\omega_r=2\pi\times 50$ Hz.  Then, for a  $^{87}$Rb
($^{23}$Na)  condensate, the healing length becomes $\xi=0.3~
(0.7)~\mu$m, the speed of sound $c=3~ (4)~{\rm mm s}^{-1}$ and the
time unit $(\xi/c)=90 ~(175)~\mu$s.

\subsection{Cross-talk of two vortices}

Each well is no longer circularly symmetric and so the vortex displacement now becomes sensitive to the direction chosen.  As such we will specify the $y$-displacement of the vortices rather than the radial displacement.  Figure \ref{fig:two_vortex_phase_diagram_split_trap} shows the phase diagram for one and two vortices in the split trap, as a function of the vortex position $y_{v}$ and the trap barrier height $V_0$.  They show close agreement with the corresponding plots for the double harmonic trap, Figs.~\ref{fig:phase_diagram_single_vortex} and \ref{fig:two_vortex_phase_diagram}(a), and demonstrate the same regimes of the vortex dynamics, e.g. cross-over and inductive dynamics.

\begin{figure}[t]
\centering
\includegraphics[width=0.9\columnwidth,clip=true]{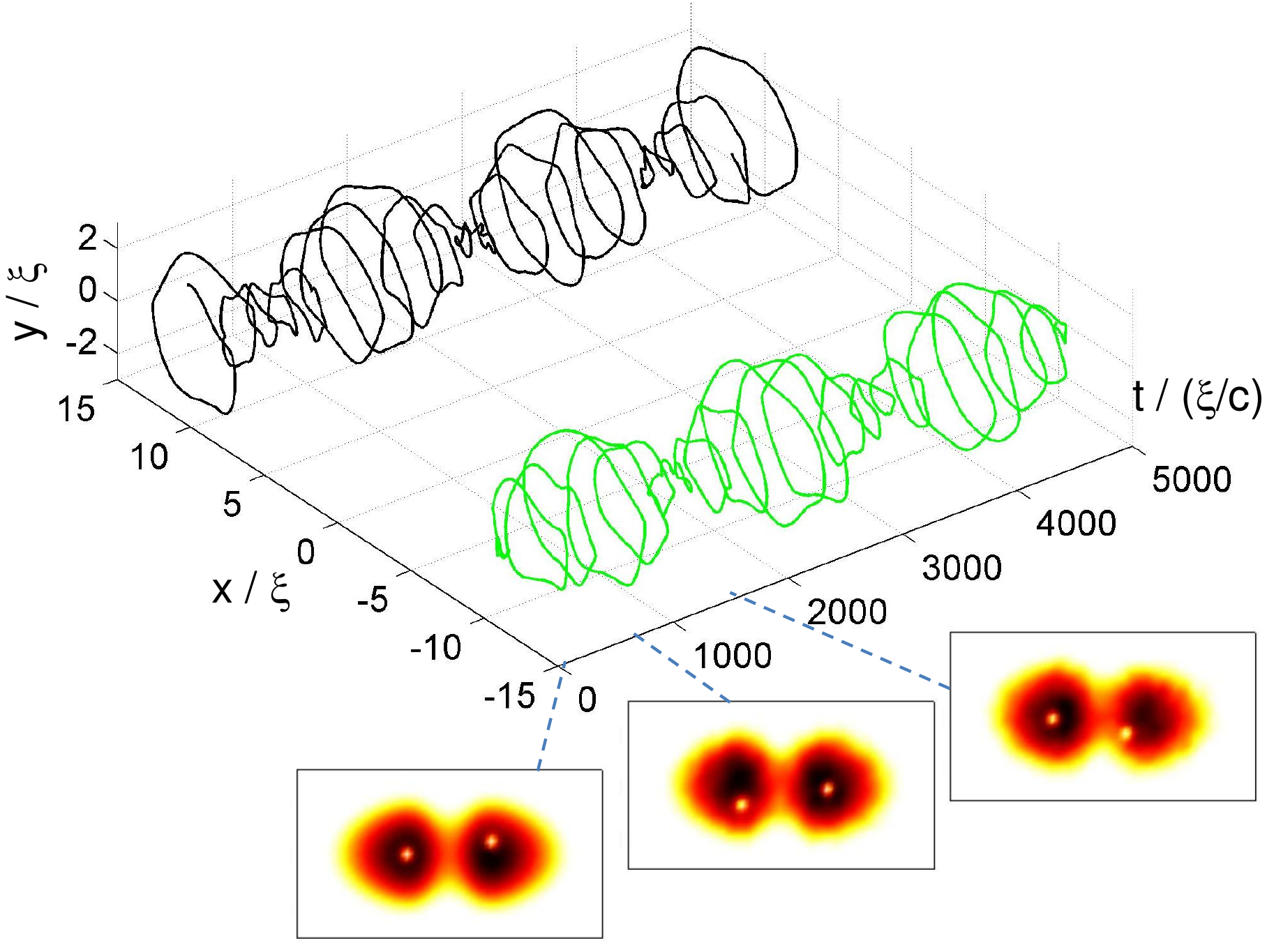}
\caption{Vortex motion during cross-talk in the experimental split trap.   Also showns are plots of the condensate density  at times $t=0$, $850$ and $1550$ ($\xi/c$). The vortices move anti-clockwise with respect to these density profiles.  The color-density mapping is defined as black/white=high/zero density.  Parameters: $V_0=0.5\mu$ and $y_{v}=3\xi$. }
\label{fig:split_trap_two_vortices}
\end{figure}

Choosing a case from the two vortex system
(Fig.~\ref{fig:two_vortex_phase_diagram_split_trap}(b)) where the
vortex dynamics is stable [$y_{v}=3\xi$  and
$V_0=0.5\mu$ ], we show the evolution of the
vortices in Fig.~\ref{fig:split_trap_two_vortices}.  The vortices
clearly drift in and out of the trap centre, out of phase with each
other.   This corresponds to the same transfer of energy observed in
the idealized trap in Fig.~\ref{fig:double_vortex_radii}(a).  The
period of the energy transfer is around $1600 (\xi/c)$ which is around 8 vortex precessions.
Snapshots of the condensate density show the initial appearance of
the condensate (vortex 1 at large radius and vortex 2 at the trap
centre), after a half-cycle of energy exchange (vortex 1 at the trap
centre and vortex 2 at large radius) and after a full cycle of
energy exchange (vortex 1 returns to large radius and vortex 2
returns to the trap centre).

\subsection{Precessing obstacle}

\begin{figure}[t]
\centering
\includegraphics[width=0.85\columnwidth,clip=true]{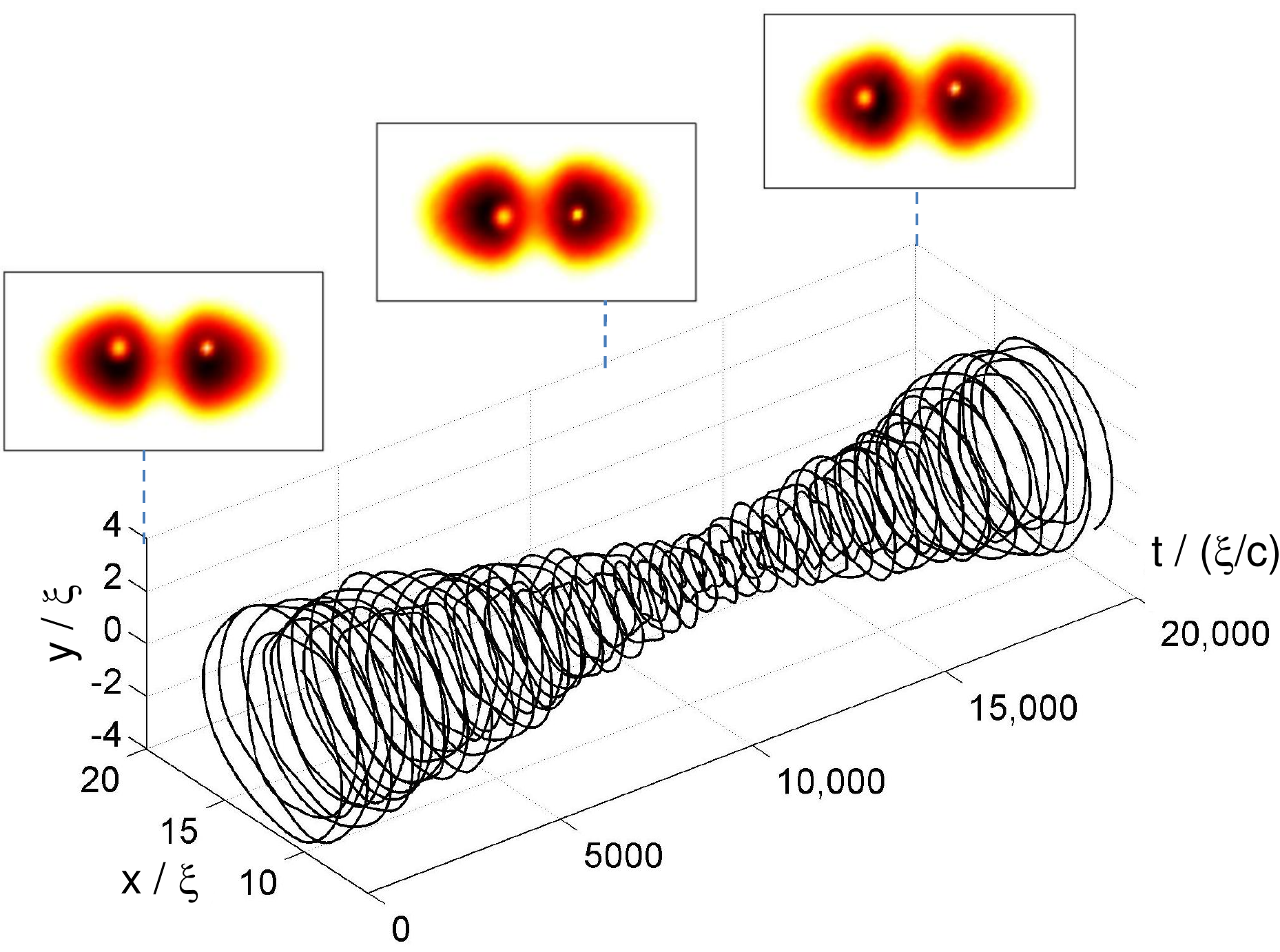}
\caption{Vortex motion under parametric driving by a precessing obstacle in the experimental split trap system. The condensate density is presented at times $t=0$, $12000$ and $20000$ ($\xi/c$).  In the density plots, the hole in the left-hand well corresponds to the obstacle and the hole in the right-hand well to the vortex, and both move anti-clockwise.  Parameters: $y_v(t=0)=3\xi$, $V_0=0.5\mu$, $\sigma=2\xi$, $A_{\rm ob}=\mu$, $y_{\rm ob}=3\xi$ and $\omega_{\rm ob}=0.234 \omega$). }
\label{fig:split_trap_obstacle}
\end{figure}

We now turn to the system with one vortex and a precessing obstacle.   The addition of a precessing Gaussian obstacle to a harmonic trap has been demonstrated experimentally \cite{Raman2001,Neely2010}.   The obstacle parameters we employ are obstacle width $\sigma=2\xi$, obstacle amplitude $A_{\rm ob}=\mu$, $y_{\rm ob}=3\xi$, $\phi_{\rm ob}=0$ and $\omega_{\rm ob}=0.234 \omega$. The ensuing dynamics of the  vortex is shown in Fig.~\ref{fig:split_trap_obstacle}. The vortex gradually spirals into the trap centre, indicating the steady increase in its energy and angular momentum.  As in the double harmonic trap, the vortex-obstacle energy transfer is slower than the vortex-vortex case, with minimal position/maximal energy reached after $\sim 10^4 (\xi/c)$.
Note that the split trap system demonstrates the same resonant behaviour as in Fig.~\ref{fig:moving_obstacle_resonance}, for example, there is a sharp resonance in the obstacle precession frequency which closely coincides with the precession frequency of the unperturbed vortex ($\sim 0.234\omega$).

\subsection{Experimental observation}

The vortex-vortex and vortex-obstacle energy transfer could be
observed through the evolution of the vortex position as it spirals
in and out of the trap.  Real-time tracking of vortices was recently
demonstrated \cite{Freilich2010}. In brief this involves
transferring (via pulsed microwave radiation) a small proportion of
the BEC into an untrapped state, allowing this representative cloud
to expand such that the vortex cores become optically resolvable,
and performing absorption imaging of the cloud. By repeating this at
time intervals, the vortex motion can be tracked.  For $^{87}$Rb
($^{23}$Na) the period of the energy exchange for the vortex-vortex
case presented here is $\sim 0.15 ~(0.3)$s, while for the
vortex-obstacle case it is $\sim 1.8~(3.5)$s.  These timescales are
well within the experimental lifetime of vortices in BECs, which can
be as large as $10$ s \cite{Rosenbusch2002}.

\section{Conclusion}

We have shown that two vortices in a double trap can undergo a
coherent cross-talk, periodically exchanging energy and angular
momentum over long-range via sound waves. These observations are
strongly analogous to the interaction of dark solitons mediated by
linear-momentum-carrying sound waves. This adds further evidence for
the striking acoustic similarities of vortices and dark solitons,
despite no mathematically rigorous link.  Sound waves artificially
generated by a moving obstacle are similarly found to drive energy
into a vortex. Crucially, for the sound to be absorbed by the
vortex, it must carry angular momentum of the same sign as the
vortex; if the sound has opposite angular momentum it is essentially
invisible to the vortex.

Our observations are robust: we have confirmed that we observe the
same qualitative dynamics for different condensate parameters (e.g.
a large, more Thomas-Fermi condensate) and double trap parameters.
Importantly, our observations occur in experimentally realizable
double well geometries, and within both the timescales of current
vortex experiments and the timescale of other dissipation
mechanisms, e.g. thermal dissipation
\cite{Fedichev1999,Jackson2007}.  Our prediction of parametric
driving of a vortex using a precessing obstacle could be used to
counteract thermal dissipation of vortices, as suggested for dark
solitons \cite{Proukakis2004}.

Trapping appears to promote sound absorption, suggesting that this
may play a much greater role in trapped BECs than in homogeneous
superfluids.  An example may lie in the formation of vortex lattices
in rotating atomic BECs.  The nucleated vortices are observed
experimentally to crystallize into a vortex lattice, a process that
requires dissipation.  While this is provided by thermal dissipation
at raised temperatures \cite{Lobo2004,Penckwitt2002,Tsubota2002},
experimental and theoretical observations at very low/zero
temperature suggest that crystallization is temperature-independent
\cite{Hodby2001,Abo-Shaeer2002,Parker2005,Parker2006}. Vortex-sound
interactions may provide such a mechanism.

Establishing the rudimentary interactions between vortices and sound
in quantum fluids will aid in exploring and understanding more
complex scenarios, such as those present in large-scale quantum
turbulence \cite{Vinen2010} and analogs of black hole superradiance
based on scattering of sound waves from superfluid vortices
\cite{Slatyer2005, Federici2006}.  In future work we hope to explore
sound absorption in 3D vortex lines and pursue acoustic ray models
of sound absorption in trapped condensates.

We acknowledge funding from the EPSRC.

\end{document}